# Inter quark mass split and a proposed possible 5 quark state

## K. W. Wong[1], G. Dreschhoff[1], H. Jungner[2]

[1]Department of Physics and Astronomy, University of Kansas, Lawrence, Kansas, USA
[2]Dating Laboratory, University of Helsinki, Finland
Email: kww88ng@gmail.com, giselad@ku.edu, hogne.jungner@helsinki.fi

**ABSTRACT**

The calculation of the meson masses based on the 5D homogeneous space-time quantum projection was shown explicitly. There are no adjustable parameters, except the rest quark mass. The discovered p-p 125 GeV resonance was proposed by us as a six quark state. Thus in this article, we like to propose a 5 quarks yet unreported baryon based on a uudu*d state, which should have a mass of approximately 33.6 GeV.

## 1. Introduction

In our previous papers in arXiv, and JMP [1a,b,c] we gave a model on the electromagnetic splitting of the $\pi$ mesons, both from a static model and a dynamic model based on Bohr-Sommerfeld orbital quantization. Since then, two years later, there was new data published on the J/$\psi$ meson [2] predicted in our article. Furthermore this resonance has been interpreted as the experimental verification of the existence of the Higgs condensed Bosonic ground state in place of the Lorentz space-time vacuum. [3]. We believe a composite p-J/$\psi$ baryon resonance of 125 GeV also exist on top of the 125 GeV boson resonance created by p-p binding. We therefore feel that a more detailed paper on the mass splitting is called for, as there are no adjustable parameters, including relativistic effects, based on our homogeneous 5D space-time projection grand unified theory [4]. Before going into the calculations, we refer our readers to references [1a,b,c and 4] for a general review of the theory. The projection breaks the homogeneous 5D space-time symmetry into SU(2)×L and SU(3)×L, where L is the 4D Lorentz space-time. Due to quantization and gauge invariance, the generators for the SU(3) group is given by the Gell-Mann quarks standard model, which consists of 3 fractional charges; namely (2/3)e, (2/3)e, and (-1/3)e. We distinguish the two different 2/3 charge quarks with the ( )' and ( ) label as it will come in easier to distinguish between different meson states when we deal with the SU(3) representations. In this paper we shall show in detail how the meson and baryon masses are uniquely determined. Since no fractional charge states are observable in L, due to gauge invariance, the distinct quarks require 3 independent data input from the meson or baryon masses to completely determine the masses of all of the rest with one single parameter, that of the bare quark mass.

Before the detailed presentation is given, we like to review how the SU(3) meson representations are obtained from the theory. The lowest mass mesons are composite two quarks states, and via gauge invariance, via the unit flux quantum of h/e, their representation is therefore given by (1) x (8). The (8) representation is an octet. However, the lowest gluon repulsive potentials for mesons between the pair of quarks that conserve gauge is generated by intermediate pair quark currents. Therefore these gluon potentials have strengths proportional to the product of the square of the quark charges from intermediate quark currents and not to the meson quark quantum signatures [4]. Such gluon potential strengths then can be labelled by the 3 colors, and 3 flavors. As the gluon potential is proportional to $s_i^2 \times s_j^2$, where $s_i$ represents the 3 fractions from the intermediate quark currents. Hence there are only 3 distinct strengths, namely $(1/3)^2(1/3)^2$, $(1/3)^2(2/3)^2$ and $(2/3)^2(2/3)^2$. By proportion they are 1, 4, 16 relative to each other. Due to gauge invariance constraint, the gluon potential then generates a mass for the composite quarks in the different hadrons. Thus for the mesons, within the pion-Kaon octet only 2 gluon potentials are involved. To cover all the gluon potentials, the octet representation is split into two. To show how to calculate for each hadron mass, we shall study the mesons within the lightest pion-Kaon octet as a detailed example of our mass calculations. Hence we will only deal with the pure 1 and 4 relative strengths gluon, represented by the pion-Kaon octet. It is simple to observe that these 2 gluon potentials give rise to this lowest meson octet, containing the pions, eta, and the Kaons. The remaining gluon potentials with strength factors 4 and 16 generate the other meson octet. There are a total of 4 such strength factors term each. Adding the single strength factor 1 term, we have exactly 9 total gluons, labelled by the product of 3 colors and 3 flavors. For the lowest mass octet, it is given by the 1 and 4 strength gluon potentials, that is the pion- Kaon octet. The Kaons and eta mesons are roughly 4 times heavier than the pions. Thus it is abundantly clear that the repulsive gluon potentials must be contributing the major portion to the individual meson mass, while the inter-quark electromagnetic interactions provide the refined mass splittings.

The purpose of this paper, therefore is not just to illustrate how the hadron masses can be computed based on our Projection-Gell-Mann standard model, but to suggest experiments to separate between this model and the Higgs model.

## 2. General meson mass formulation

As mentioned earlier, we need 3 independent distinct data inputs. We shall choose those from the charged pions, the neutral pion, and the eta meson. Note the $\pi^+$ and $\pi^-$ has identical mass, they are not anti-particles of each other. Thus $\pi^+$ is the combination of (2/3) and (-1/3)*, while $\pi^-$ is from (2/3)* and (-1/3). The * denotes the anti-particle. The masses of the $\pi$ mesons as mentioned originate from two parts. The majority comes from the quantum gauge confinement on the repulsive gluon potential, which is the product of two vector potentials generated by the intermediate quark currents. Since the $\pi$ mesons are the lightest mesons, this gluon potential is therefore generated by the intermediate quarks of 1/3e, and -1/3e. The explicit form of this gluon potential was published earlier [1c]. The exact value of the gluon generated mass component depends on two important variables. Namely the loop radius from the gauge confinement and the velocity, and thus the currents of the intermediate quarks. Both of these two parameters are not arbitrary and can be determined through the detailed dynamic model. We shall show the steps below.

Let us assume the gluon generated mass as $M_o$, which is due to the gluon of charge generated strength of $(1/3)^2 \times (1/3)^2$ for the pions. Note that knowing the strength factor alone is insufficient to quantify the gluon strength, as it also depends on the intermediate quark velocities, which is the result of the conformal projection [see ref.4]. Then the electromagnetic mass splitting between the charged and neutral pion comes from the two quarks interaction as must be given by their Hamiltonian differential Schrödinger eigenvalue solution. For the charged pions, they are formed by the pair, with charges (-1/3)*(2/3) and (-1/3)(2/3)'*. It should be noted although $\pi^+$ is not the anti-particle of $\pi^-$, as they contain the two separate different (2/3)e quarks, their electromagnetic interactions are the same, their masses must be identical. The neutral pion as a charge 0 state is composed of $(-1/3)$e and (-1/3)*e quark pair. With such a choice, the eta meson, having the same quantum signature is also formed by the (-1/3)e, (-1/3)*e pair. There remains two other neutral particles, that of (2/3) (2/3)'* and (2/3)'(2/3)*, which are the two neutral Kaons in the octet representation. Its mass can be calculated for confirmation of our theory.

The eigen energy produced by the quark pairs of the Schrödinger equation consist of two parts due to separation of coordinate variables of their motion in the center of mass coordinate frame, where the center of mass can be treated as stationary. From conformal projection, to preserve gauge invariance, the charge per unit mass quantity $\alpha$ is an invariant constant (see ref. [4] for detail.) Therefore, for a pair of quarks $s_i$, $s_j$, we have the center of mass $m_c$

$$m_c = (|s_i| + |s_j|)m. \tag{2.1}$$

where m the rest mass of the bare quark, is a universal constant, and can only be determined by data fitting. Once found, all other mass generated interaction masses are calculated.

The reduced rest mass is then given by $m_{i,j}*$

$$m_{i,j}* = \{|s_i s_j|/(|s_i| + |s_j|)\}m. \tag{2.2}$$

Hence, for the $\pi^+$ and $\pi^-$, their $m_c$ and $m*$ are the same.

Because of the quark charges within each meson, they will produce a Coulomb potential

$$V_{i,j} = s_i s_j e^2 / r_{i,j} \tag{2.3}$$

This potential energy can be either repulsive or attractive depending on the sign of $s_i s_j$.

Because of quantum mechanics the separation distance $r_{i,j}$ must obey Bohr-Sommerfeld quantization depending on the orbital dimension constraint. For example for a 3D model for hydrogen problem, the Bohr-Sommerfeld quantum is given by $L = nh/2\pi$, with the ground state orbit defined by $n=1$. However for the 2D non-relativistic hydrogen, the energy spectrum is $-R/(n+1/2)^2$, where R is the Rydberg constant, and the ground state is given by $n=0$, thus the lowest orbital radius is only half of the 3D model ground state value, (see section 6 for the exact solution of the relativistic 2D hydrogen, which gives the same) and therefore dependent on the pair's relative velocity. The ground state radius is given by

$$r_{i,j} = h\gamma/2\times 2\pi \; m_{i,j}^* v = h\gamma/4\pi \; [c \; m_{i,j}^* \; (1-\gamma^2)^{0.5}]. \qquad (2.4).$$

where the relativistic factor $\gamma = [1-(v/c)^2]^{0.5}$, and $m^*/\gamma$ is the kinetic energy.

Substituting $r_{i,j}$ from eq.(2.4) into $V_{i,j}$ in eq.(2.3), we get the Coulomb potential energy

$$V_{i,j} = 4\pi s_i s_j e^2 \; c \; m_{i,j}^* \; (1-\gamma^2)^{0.5}/h\gamma. \qquad (2.5)$$

Including both the kinetic and potential energies for the $\pi^+$ and $\pi^-$, we see that their inter-quarks Schrödinger equation is identical. Hence, their masses are the same. Instead of solving immediately for the solution, let us simply assume a total mass value of C for a meson state composed of quark pairs. Apart from the gluon contribution we can study the total energy T.E. of the quark pair within the meson.

$$T.E. = m_c + m^*/\gamma + V. \qquad (2.6)$$

Note, all 3 terms in the T.E. must be the total sum over all the quark pairs within the meson state, as the neutral mesons contain two different opposite charged pairs.

Before trying to fit data and determine both the quark mass m and the relativistic factor $\gamma$, it is interesting to compare the result given by eq.(2.6) to that obtained from the exact Chern-Simons solution, given in ref.[5], because there $\gamma$ is calculated. Therefore the ground state of the hydrogen, the binding energy is equal to the reduced mass $m^*$. See ref. [5], eq. (3.19), which requires angular momentum $L+ s_z + \Delta = \overline{\alpha}^2$, where $\overline{\alpha} = e^2$. By choosing the Chern-Simons associate flux $\Delta = 0$, then the ground state corresponds to reducing to simply substituting Bohr-Sommerfeld quantized orbital radius for the Coulomb potential by setting $m^*+V=0$. This exact solution is not a semion state. From that we easily find $\gamma=0.18$ (see section 6). Thus T.E. for $\pi^0$ then reduces to

$$T.E.= m_c + m^*[1-(1/9)\gamma]/\gamma \qquad (2.7)$$

For pure curiosity, should we assume a hydrogen composed of a negative charge having a reduced mass $m^*$ equal that of the bare quark mass, and $m_c$ is stationary, then this object carries an effective relativistic mass energy $m[1-\gamma]/\gamma$, which would give the minimum possible T.E. value.

## 3. The gluon potential

The gluon potential obtained from the projection theory is given by Eq. 8.12 of ref. (3), namely

$$U = \sum_{\mu\mu'} \frac{e^2 m^2 v_\mu v_{\mu'}}{(2\pi)^2} \times (\alpha\Phi_0)^2 s_i^2 s_j^2 / r_0^2 \qquad (3.1)$$

where $\mu$, $\mu'$ runs from 1 to 3, that represent the 3D vector components of the quark i and j, while $r_0$ is the radius of the unit quantum flux, and not that of the relative distance between the two quarks within the meson. Here $\alpha$ is a constant given by charge per unit mass, as defined in ref [4]. This expression of the gluon potential generated by the intermediate quark currents product shows it is dependent on the total quark rest mass, and the velocity $v^2$, as expressed in the center of mass frame, summing $v_\mu v_{\mu'} = v^2$, and is not dependent on the individual charge fractions of the quarks within the meson state, which came from the projection, and also implies that $v^2$ of all quarks are equal, if all hadrons within a SU(3) representation appear to have the same gluon generated mass component. A requirement on the conformal projection and the resulting fractional charge quarks correlation to the vector potential fields, allowing the gauge invariance to be independent of the quark fractional mass, imposed by the Lorentz energy- momentum metric.

As such knowing the relativistic factor $\gamma$ of all quarks that generates mesons will also determine the gluon strength multiplier constant. Hence the two basic parameters required on the conformal projection action, implies this projection action requires a total energy $mc^2/\gamma$, consistent with the space-time conformal projection, and uncertainty principle, that a finite fixed proper time amplitude for masses is required.

Thus eq.(3.1) becomes

$$e^2 m^2 (1-\gamma^2)/(2\pi)^2 \times (\alpha.\Phi_0)^2 s_i^2 s_j^2 / r_0^2 \qquad (3.2)$$

in mass unit.

Hence the gluon generated mass can be expressed as n×$M_o$, where n is 1, 4 and 16, with

$$M_o = e^2 m^2 (1-\gamma^2)/(9 \times 2\pi)^2 (\alpha \times \Phi_0)/r_0^2 \qquad (3.3)$$

This gluon potential contributes the predominant amount of mass to the mesons. It is important to analyse further the gauge confinement. Since $M_o$ is formed by the product of the vector potentials generated by the two intermediate quark currents having charges of $(1/3)e$, and $–(1/3)e$, each $(1/3)e$ charge produces a magnetic field perpendicular to their loop current. If these two current loops are aligned, then this H field would be very strong as the loop radius is very small. However, because the + and - charges produce opposite currents, such that the quantum flux unit is h/e, the H field generated cancels. Should these 2 currents be in phase, the flux quanta will be h/2e, and the gauge confinement radius would be halved, leading to 4 times greater gluon generated mass. Eq. (3.3) can be rewritten in MKS units. We get

$$M_o = \mu_0 e^2 (1-\gamma^2)(h/2\pi)^2/[9c^2(h/2\pi)^2 r_0^2] \qquad (3.4)$$

where $\mu_0$ is the permeability constant.

Leading to (see section 6 for more detail)

$$r_0 = 2.728 \times 10^{-16} \text{ m} = 2.728 \times 10^{-3} \text{ fermi.} \qquad (3.5)$$

The fixing of $r_0$, the quantum gauge loop confinement parameter, can be compared to the Bohr-Sommerfeld radius of the collapsed ground state, which is derived from the exact relativistic 2D hydrogen solution [5], where the exact eigenvalues are given by the vanished associate flux, as shown in eq. (3.4) of Ch. 3 of reference [4]. We obtained from the Bohr Sommerfeld quantization, the 2D Bohr radius for $\pi^0$, $R_o = 3.127 \times 10^{-2}$ fermi, which is larger than $r_0$ the gauge parameter. Should we change the quantum flux to h/2e, the gauge loop radius will decrease by half to $1.364 \times 10^{-3}$ fermi, only 1/20 that of that of the G.S. Bohr radius. Thus this allows the p-p 125 GeV composed of 6 quarks to exist.

If without spin and $L = e^2/(2\pi\mu^2)$, where $\mu$ is the Chern-Simons term coupling, or the spin cancels the Chern-Simons associate flux $\Delta$, then $R_0$ vanishes, leading to a semion state. The Coulomb potential then is singular, unless the quark charges vanishes. In short, we do not allow the existence of quarks inside the Bohr limit. Such singular solutions is an inherent property of -1/r potentials in 3D as well as in gravitation. The other finite $R_0$ solution reproduces the Bohr-Sommerfeld orbit and preserves statistics. Therefore, the 2D hydrogen ground state of finite $R_0$ reveals that the singularity is removed by a finite void core space exclusion of the solution, which we found from the 5D projection theory [4]. This physical reason for treating the evaluation of $R_0$ as the 2D relativistic hydrogen comes from the extra physics that the ground state binding energy of this 2D hydrogen is equal to the rest mass of the electron, making the total energy of this orbiting electron exactly zero and preserves the notion of creation from nothing, not dissimilar to the Higgs theory. In the 5D projection quantum theory, masses and charges are created out of nothing due to dimension reduction, which means any projection created eigenstate must satisfy the conservation of binding energy to mass, which the 2D solution obtained from 3D space dimension reduction also satisfies. Since $r_0$ is a gauge confinement parameter which is invariant for all allowed states, we can select the meson state of charge +e, and the intermediate current state as that of -e then this combined system can be viewed at as a hydrogen model. By further viewing the +e charge as at the center of mass, and treated as stationary, then we can obtain the Chern-Simons 2D model, for the derivation of $r_0$. However, this method when applied to the baryons need corrections due to the 3 body problem. Thus even without getting into the semion state, it could lead to $R_0$ splitting, particularly for the lowest octet massive objects. This $R_0$ splitting probably is revealed from the small mass difference between $\Sigma^0$ and $\Lambda^0$ at the center of the proton, neutron octet, as their quark quantum signatures are presumed identical, and hence the inter-quarks total energies should be equal, without correction due to Chern-Simons coupling $\mu$ modification on the associate flux from the 3 body correction. It is, however, not difficult to see the origin of this three body effect. To obtain the zero charge requires the sum of 2/3 to two -1/3 charges, or vice versa. Expansion in pair representations gives however two different pairs final forms of 1/3, -1/3; or 2/3, -2/3, depending on the order of pair iteration for the 3 body coordinates. Hence in terms of a hydrogen model, the resulting Coulomb strengths could be different, and with the Chern-Simons coupling $\mu$ could lead to an order created modification. Such order fine splittings of $R_0$ probably should only be observable for the lowest 2 strengths as it is independent of the gluon strength factor $s_i^2 s_j^2 s_k^2$. The 3 body correction to the 2D Chern-Simons hydrogen solution is a difficult problem and remains to be solved [5].

Coming back to the evaluation of the meson mass these solutions for $r_0$ and $R_o$ are substituted into the general quadratic meson mass equation, we get

$$M^{*2} = (nM_o)^2 + (m_c + K.E. + V)^2 \qquad (3.6)$$

where K.E. and V are also expressed in mass unit.

## 4. Determination of the relativistic factor $\gamma$

With the meson mass formula given by eq. (3.4) and eq. (3.6) we can now proceed to determine the relativistic factor $\gamma$ of all quarks within a meson without assuming that we have a 2D ground state data fit. To show that, let us consider the mass difference between the $\pi^+$ and the $\pi^0$, for which the gluon contributed mass is Mo, for both, we get

$$M^*(\pi^+)^2 - M^*(\pi^0)^2 = (m_c + K.E.^+ + V^+)^2 - (m_c^0 + K.E.^0 + V^0)^2. \quad (4.1)$$

The left hand side is positive, while the right hand side can be factorized into

$$\{m_c^+ + m_c^0 + K.E.^+ + K.E.^0 + V^+ + V^0\} \times \{m_c^+ - m_c^0 + K.E.^+ - K.E.^0 + V^+ - V^0\} > 0. \quad (4.2)$$

Since $V^0$ is negative, if it is greater than $\{m_c^+ + m_c^0 + K.E.^+ + K.E.^0 + V^+\}$, then the term

$\{m_c^+ - m_c^0 + K.E.^+ - K.E.^0 + V^+ - V^0\} < 0$. In another word

$$K.E.^0 + m_c^0 > K.E.^+ + m_c^+ + V^+ - V^0 \qquad (4.3)$$

where all terms are positive.

Similarly, if $V^0$ is less than $\{m_c^+ + m_c^0 + K.E.^+ + K.E.^0 + V^+\}$, then

$$\{m_c^+ + K.E.^+ + V^+ - V^0\} > m_c^0 + K.E.^0 \qquad (4.4)$$

Equations (4.3) and (4.4) establish the limits for a solution.

With the numerical numbers given by the pion masses, the solution lays close to the limit given by (4.4).

After some algebraic manipulations, we get

$$[(1/3) + (2/(9\gamma^+) - 1/(6\gamma^0) + e^2/(mc^2 R_o) \times (1/9)] \times [(5/3) + (2/(9\gamma^+) + 1/(6\gamma^0) + e^2/(mc^2 R_o) \times (1/3)] \quad (4.5)$$

It is easy to see that the non $e^2$ terms are larger, and together with the 2D G.S. minimum limit, we have

$$\gamma^+ > (4/3)\gamma^0 \qquad (4.6)$$

In fact from pion mass data fits we found $\gamma^+/\gamma^0$ is close to 4/3, which we will verify in section 5. This result clearly shows the quarks are extremely relativistic. The single mass difference equation between $\pi^+$ and $\pi^0$, cannot determine the exact value of the bare quark mass m and the relativistic factor $\gamma$ uniquely. We require another data input, for which we will choose the $\eta$ mass.

## 5. Simultaneous data determination of m and $\gamma$

The gluon potential contribution to the eta meson is 4 times that of the pions. We shall consider

$$(4M^*(\pi^0))^2 - M^*(\eta)^2 = [4(m_c^0 + K.E.^0 + V^0)]^2 - [m_c(\eta) + K.E.(\eta) + V(\eta)]^2 \qquad (5.1)$$

Again, the right hand side can be factorized as

$$\{4(m_c^0 + K.E.^0 + V^0) + m_c(\eta) + K.E.(\eta) + V(\eta)\} \times \{4(m_c^0 + K.E.^0 + V^0) - m_c(\eta) - K.E.(\eta) - V(\eta)\} > 0.$$
$$(5.2)$$

Both $V^0$ and $V(\eta)$ are negative, and the same when cross interactions between the meson quarks constituents and the gluon potential field is ignored. In fact due to the same composition of quarks, as they belong to the exactly same quantum signatures, the total internal quark energies are then the same, thus do not impose new limits, other than $m_c{}^0 + K.E.^0 + V^0 > 0$.

We get from mass data for $4\pi^0$ and $\eta$, $m_c{}^0 + K.E.^0 + V^0 = 45$ MeV.

With equations (4.1) and (5.1), and with the knowledge of $\gamma$ limits, we find from subtracting $\pi^0$ mass from $\eta$ mass together with eq. (5.1) and eq. (4.1) numerically, $\gamma^0 = 0.2$, which is very close to the value 0.18 obtained from the relativistic 2D hydrogen ground state reported before [1b], and m=34 MeV, plus $M_o$=121 MeV as was reported before.

To check these numerical solutions, we substitute them into

$$M^*(\pi^+) = \{M_o{}^2 + (m_c{}^+ + K.E.^+ + V^+)^2\}^{0.5} \qquad (5.3)$$

where $m_c{}^+ = m$, and

$$K.E.^+ = (2/9)m/\gamma^+ \qquad (5.4)$$

Assuming the G.S. radius $R_o'$ is given by eq.(2.4), then we get

$$V^+ = (2/9)e^2/R_o = (2/9) \times (1/6)m \qquad (5.5)$$

the factor $(1/6)m$ came from the Chern-Simons G.S. to the attractive $V^0$ potential. Although $V^+$ does not lead to binding, the reduced mass from the quark pair of $\pi^+$ must still satisfy the Bohr-Sommerfeld orbital quantization, same as the pair of $\pi^0$. Which lead to $(1-(\gamma^+)^2)/(\gamma^+)^2 \times (2/9)^2 = (1-(\gamma^0)^2)/(\gamma^0)^2 \times (1/6)^2$. This equality gives $\gamma^+ = 0.237$, slightly less than $\gamma^+/\gamma^0 = 4/3$, as we obtained earlier in section 3.

As an example we can calculate $R_o$ as obtained from $\pi^0$

$$R_0 = (h/4\pi)\gamma^0/\{(1/6)m[c^2(1-(\gamma^0)^2]^{0.5}\} = 3.127 \times 10^{-2} \text{ fermi}.$$

which is far greater than $r_0$, the gauge radius, as expected. Because quarks must be in Lorentz space, while gluon fields are inside the semion radius $R_o$.

$$M^*(\pi^+) = \{121^2 + [34(1+(2/9)/\gamma^+ + (2/9)(1/6)]^2\}^{0.5} = 139 \text{ MeV}. \qquad (5.6)$$

with $\gamma^+$ given by the limit from eq.(4.6), and $\gamma^0$ =0.18, the 2D Chern-Simons solution (see section 6).

Let us further check for $M^*(\pi^0)$, we have

$$M^*(\pi^0) = \{M_o{}^2 + m_c{}^0 + K.E.^0 + V^0)^2\}^{0.5} \qquad (5.7)$$

where

$$m_c{}^0 = (2/3)m \qquad (5.8)$$

$$K.E.^0 = (1/6)m/\gamma^0 = 0.926m \qquad (5.9)$$

and

$$V^0 = -(1/9)(1/6)m \qquad (5.10)$$

Thus

$$M^*(\pi^0) = \{121^2 + [m(2/3 + (1/6)/\gamma^0 - (1/54))]^2\}^{0.5} = 132 \text{ MeV} \qquad (5.11)$$

And this also similarly gives $M^*(\eta)$=487MeV and $M^*(K^+)$=489MeV

Note that all results are just either marginally off or equal to data. Differences we believe are due to the inaccuracy in determining $\gamma$ from the mass data.

## 6. A more detailed analysis of the 2D relativistic hydrogen spectrum

The exact energy spectrum for the 2D relativistic hydrogen is given by eq.(3.19) in ref.[5] for the Chern-Simons (C.S.) hydrogen as

$$E = mc^2 \{1 + \bar{\alpha}^2/[\gamma' + n']^2\}^{-0.5} \qquad (6.1)$$

where $\bar{\alpha}$ is the Coulomb strength $e^2/\epsilon_0$, $\epsilon_0$ is the free space dielectric constant. While n'=0,1,2,3... Hence for the ground state, G.S. n'=0.

$$\gamma' = (k'^2 - \bar{\alpha}^2)^{0.5} \qquad (6.2)$$

see eq.(3.18c) of ref.(4), and

$$k' = k + \Delta \qquad (6.3)$$

k is the total angular momentum $L+s_z$, L is the quantum value of the angular momentum r×p and $s_z$ is the spin 1/2.

$$\text{while } \Delta = e^2/(2\pi\mu). \qquad (6.4)$$

see eq.(3.9) of ref.(4), μ is the Chern-Simons coupling.

It is therefore obvious that, as γ' goes to 0, the ground state E goes to 0, implying the 2D Coulomb binding collapses the system, and cancels the orbital charge's rest mass energy. It should be pointed out that for γ'=0, for the G.S. L=0, and k=$s_z$. If we set Δ equal to zero, then γ'=0, is equivalent to treating the Coulomb strength as 2 times that of $e^2/\epsilon$. Or the equivalence of summing over spin orientations.

Assuming an open boundary condition, the G.S. eigenvalue of the 2D relativistic C.S. hydrogen equation collapses. Thus satisfying

$$m^* - e^2/R_o c^2 = 0 \qquad (6.5)$$

where m* is the reduced mass.
The G.S. radius $R_o$ also satisfies 2D Bohr-Sommerfeld quantization

$$R_o = (h/4\pi) \times (\gamma/m^* v) \qquad (6.6)$$

Setting the max Q = 1 for interquark potentials we get

$$\gamma = 1/\{1 + [(h\epsilon_o c/4\pi e^2)]^2\}^{0.5} = 0.18 \qquad (6.7)$$

where $\epsilon_o = 8.85419 \times 10^{-12}$ C$^2$/Nm$^2$

For the meson states, as long as $r_o < R_o$, where $r_o$ is the gauge confinement loop parameter, the interquark potentials can be calculated by inserting γ from eq. (6.7) and $R_o$ from eq. (6.6) into the total mass as given by eq. (3.4). Because the gluon potential in mesons, given by eq. (3.2) is not as short range as that of the baryons (see eq. 8.32 in ref. [4]) such 2D hydrogen G.S. might not be possible if $r_o > R_o$ for the very massive 4 and 6 quark boson states.
To illustrate our point let us consider the proton (uud) and the neutron (ddu). Both states are a 3 body problem for which there is not yet a solution. None the less we can approximate the net coulomb potential by summing over pairs. We get for the proton

$$V_p = (2/3)(2/3)e^2/r - 2 \times (1/3)(2/3)e^2/r = 0 \qquad (6.8)$$

irrespective of the value of r provided it is the same for all pairs. This constrain can be obtained by imposing net current zero as carried by all pairs. For the proton, there are 2 pairs of (2/3)e and (-1/3)e quarks, with net charge of (1/3)e each, and one pair with (2/3)e, and (2/3)e quarks, with net charge (4/3)e. Therefore their net current is given by 2×(1/3)ev+(4/3)ev'=0.

Thus v=-2v'.

For v not zero, these pairs then must carry kinetic energy. As the total energy is minimized together with the potential energy. We obtain v=0, as well as V=0. Hence all separation distances in the pairs are equal and stationary, making the K.E. simply the sum of the pairs reduced rest masses. Therefore the lowest T.E. of these 3 quarks is

$$\text{T.E.} = [2(2/3) + (1/3)]m + [(2/3)(2/3)/(4/3)]m + 2\{(1/3)(2/3)/[(1/3)+(2/3)]\}m = (22/9)m \quad (6.9)$$

As there is no potential to create K.E.

The effective proton mass

$$M_p^2 = \overline{M_o}^2 + (22/9)^2 m^2 \quad\quad\quad\quad (6.10)$$

If $\overline{M_o}$ = 928 MeV and m = 34 MeV we get $M_p$ = 932 MeV.

On the other hand the coulomb potential for the neutron is

$$V_n = (1/3)2e^2/r - 2 \times (1/3)(2/3)e^2/r = -(1/3)e^2/r \quad\quad (6.11)$$

assuming again the same r value separation.

Within the neutron, the 2 pairs (ud) quarks carries a charge of $2 \times (1/3)e$, while the single (dd) pair carries a charge of $2 \times (-1/3)e$. Hence their motion will produce a net current $(2/3)ev + (-2/3)ev'$. The net zero current requirement, gives v=v' irrespective of the value of v. Therefore, all three pairs will move in phase like a rigid body when their pair separations are fixed. The effective Coulomb potential $V_n$, based on the fact that the separation distances between the quark pairs are fixed and equal, and therefore rigid, is a quantum well that trap all 3 quarks in a rigidly structured configuration. Thus the K.E. is not from the pairs, just like in the proton, but rather from the center of mass of the 3 quarks. Therefore

$$\text{T.E.} = m_c/\gamma + (11/18)m + V_n \quad\quad\quad\quad (6.12)$$

where $m_c = (4/3)m$, and (11/18)m comes from the rest energy of the 3 rigid pairs.

The $m_c$ orbital in the potential must satisfy Bohr-Sommerfeld quantization. For the G.S., we have

$$(m_c/\gamma)c(1-\gamma^2)^{0.5} \times R_o = h/4\pi \qu\quad\quad\quad (6.13)$$

It is easy to see that as $\gamma$ increases towards 1, $R_o$ goes to infinity, and any 1/r potential goes to zero. Thus as the Coulomb potential strength factor decrease, gamma must increase accordingly. For any strength factor less than 1, $R_o$ cannot reach the semion radius limit, and the system cannot collapse. With eq. (6.13) $\gamma$ can be determined, from the G.S. energy of the 2D hydrogen Schrodinger equation (see ref.4). Because the potential $V_n$ is weaker than the hydrogen by a factor of 1/3, it is not sufficient to collapse $m_c$ to the semion border. None the less, we can approximate this eigenvalue as $-(1/3)m_c$. Substituting this result into eq. (6.12), we get

$$\text{T.E.} = (11/18)m + (4/3)m[(1/\gamma) - (1/3)] \qu\quad\quad (6.14)$$

The relativistic factor $\gamma$ can be easily solved from the neutron mass, or the energy eigenvalue. We get

$$\gamma = 0.486. \quad\quad\quad\quad\quad\quad (6.15)$$

which is far less relativistic then the Chern-Simons G.S., where $\gamma$ is equal to 0.18. In fact it is nearly 3 times larger, confirming that the approximation of the $V_n$ eigenvalue as $-(1/3)m_c$ is quite good. Taking into account orbital quantization restriction as given by eq. (6.13), the value of $\gamma$ = 0.486 is exact, as deduced by the Chern-Simons limit. As the eigenvalue requires $\gamma/[1-\gamma^2]^{0.5} = 3 \times 0.18/[1-(0.18)^2]^{0.5}$. Thus giving $\gamma$ = 0.486. From this result, the energy eigenvalue for $V_n$ is indeed $-(1/3)m_c$. It is not an approximation as assumed earlier.

We can now compute the neutron mass from eq.(6.12), and the gluon generated mass $M_o$, which is equal to 928 MeV. Note this value is obtained from the gauge radius $r_0$ and the Chern-Simons G.S. $\gamma$ value of 0.18, and not an adjustable parameter. [see our previous discussion on the meson gluon value.] Hence we get

$$M_n^2 = 928^2 + \text{T.E.}^2 = 928^2 + 34^2[(11/18)+(4/3)(1/0.486 - 1/3)]^2. \qu\quad\quad (6.16)$$

The neutron mass is then Mn=933.27 MeV with no adjustable parameters, except for the bare quark mass of 34MeV as an input, which is nearly exactly the value given by data. The fact that the neutron relativistic factor is quite large, means both the proton and neutron are quite stable as compared to the mesons, where even for the $\pi^+$ the relativistic factor is much more relativistic, and therefore the mesons are more unstable.

As mentioned in this paper, a 5 quark state of the lowest gluon potential will generate a mass of 33.6 GeV. To show that, we follow our previous approach in estimating a 6 quark state starting from 125 GeV that was found by CERN experiment. The lowest 5 quark state would be products of 5 vector potentials generated by 5 intermediate quark currents. For the lowest possible strength, we will select them as products of the gluon potentials that generate the pion and the proton. As such, we treat it as a second order perturbation. Let $U_2$ be the gluon potential for generating the pion, and $U_3$ that of the proton, then in second order perturbation, we have $U_5 = U_2|2>(1/E_2)<2|U_3 + U_3|3>(1/E_3)<3|U_2$, where $|2>$ represents a 2 quark intermediate state, and $|3>$ that of a 3 quark state. $E_2$ then is the gluon generated mass energy to the pion, while $E_3$ that for the proton. Thus this 5 quark gluon must obey h/2e flux quantization instead of h/e, otherwise they will simply become direct products. In fact all 4, 5, 6 quark states obey flux quantization gauge invariance quantum constraint of h/2e, whereas the standard model unit flux obeys h/e. The r dependence of this 5 quark gluon potential is obviously $1/r^5$. Due to the reduced flux quantum, the gauge confinement would produce an increase in value of $2^5$ as compared to the decoupled $U_2 \times U_3$. For the pion, we have $E_2 = 121$ MeV, and for the proton $E_3 = 928$ MeV as we reported [1c, 4].
Thus substituting these numbers, we get

$$<5|U_5|5> = 2^5\{928+121\} \text{ MeV} = 33.6 \text{ GeV} \qquad (6.17)$$

Like the p-p 6 quark state, where the gluon potential thus obtained is 119 GeV with the remaining 6 GeV mass coming from the inter quark total energies. Hence, depending on the signatures of the 5 quarks, such a state would have a total mass some 2 to 4 % greater than 33.6 GeV.
Should we change the $U_2$ to that, that generates the J/Ψ singlet, this 5 quark mass would be much larger, as we would replace the 121 MeV pion to that 3021 MeV from the J/Ψ, and get $2^5(928+3021)$MeV = 126.4 GeV. Therefore finding a 125 GeV resonance need not mean it has to come from a condensed universal Higgs bosons ground state.

## 7. Conclusion

In the pion-Kaon octet only gluon strengths of 1 and 4 are used in their mass generation. The much larger strength 16 must then create heavier mesons. It is important to realize that the gluon strength factors must satisfy a sum rule as discussed in detail formerly in ref. [1c and 4], which led to regrouping of the different gluon strength potentials. It is this regrouping that leads to the identification of the singlet as the J/Ψ, with the gluon potential $2(2/9)^2+\sqrt{2}[(4/9)^2-(2/9)^2]$. What we like to point out here is not to repeat the details given in the previous papers, rather to mention that apart from that discussed, the need to regroup the terms in the sum rule is also revealed in the neutral π and η meson present at the center of the pion-Kaon octet. Note first, it comes from the gluon strength potential exactly the same as that of the Kaons. However, it is composed of a neutral particle formed by $(2/3)(2/3)^*+(2/3)(2/3)^{'*}$. The intermediate currents generated by such an intermediate quark pair state, would belong to the missing strength term of 16. Hence it is easy to see how the gluon strength sum rule must be regrouped as discussed before in ref. [1b and 4]. The remaining point, we want to stress here, is that the gluon mass contribution to the J/Ψ, due to the value of $M_o$ found from the pion data was reported [1a] as 120 MeV. Hence the J/Ψ mass is of value $\{2(2/9)^2+\sqrt{2}[(4/9)^2-(2/9)^2]\}M_o=24.98M_o= 2996$MeV. The J/Ψ mass is 3096 MeV. This difference of 100 MeV represents a 3% error, and cannot be completely covered by the inter-quark Coulomb correction. Actually, it is easy to get this Coulomb correction numerically, which would only be able to add roughly 1% to the final mass. Therefore, this remaining 2% correction actually could be from the error made in the γ factor =0.2. A slightly increased value of 1-γ, from 0.8 as given in our previous paper to 0.82 as obtained from the 2D G.S. here, leads to a slight increase in $M_o$ from 120 MeV to 121 MeV, while the quark mass remains unchanged at 34 MeV. This change increases the J/Ψ gluon contributed mass from 2996 MeV to 968+2053 = 3021MeV. The singlet J/Ψ is composed of all gauge invariant 9 pairs, namely (1/3)(-1/3); 2×(1/3)(2/3); 2×(-1/3)(-2/3); and 4×(2/3)(-2/3). Thus its $m_c$=10m, and its reduced mass M* is $[(1/6)+4(2/9)+4(1/3)]$m=(43/18)m. The Coulomb potential V=$e^2\{-(1/9)+4×(2/9)-4×(4/9)\}$/r=-$e^2$/r. This negative potential would lead to the Chern-Simons G.S. Thus V = -M*. Now substitute these results into the J/Ψ mass formula, we get a final mass value of exactly 3096MeV by using γ equal to 0.19 instead of 0.18, the Chern-Simons semion limit. This is reasonable, because in the reduced masses, we have 9 terms. Not all of them can be in this Chern-Simons G.S. orbital. It is expected this orbital would be somewhat broadened leading to an averaged value around γ = 0.19.
Perhaps more interesting is not the perfect numerical fit, but rather that the 1-γ value of 0.8 gives us γ=0.2. It is interesting to investigate how we can create such a 2D hydrogen from the 5D projection. It was shown, that the conformal projection can create quarks that through gauge confinement would produce a proton, with a rest mass of nearly 1 GeV. Should this proton be assumed as that heavy charge that would bind through 2D Coulomb force, with a negative charge of -e, composed of two quarks, with charges -2/3e, and -1/3e, which also has a combined rest mass of at least 34MeV, then this neutral atom like particle G.S. has the equivalent reduced mass higher than 102MeV. This state is of course not the muon. The combined neutral particle of $\pi^-$ and p is a spinor composed of 5 quarks with a mass greater than 1.042 GeV, the total mass from just adding M($\pi^-$) and M(p), and if it actually exists has yet to be found? In fact if this 5 quarks state of uudu*d can be modelled as a 2D hydrogen state of a proton and a $\pi^-$, and if we derive this 5 quark gluon potential by 2nd order perturbation, we would get a gluon mass component of 33.6 GeV (see Section 6 for detail). If this state exists, it probably would quickly decay into a neutron and a stream of mesons? - as that would lower the mass energy.

It should be pointed out that all gluon generated bound quark states beyond two and three quarks, their quantum gauge confinement is based on the quantum flux of h/2e instead of h/e, and would therefore be much more massive. What remains most important for our grand unified theory is to totally reformulate the Perelman mapping [6a,b] in a quantized covariant form, such that one can clearly understand the quantum correspondence of Poincaré Conjecture, for which it is needed to reformulate the Lorentz-Riemannian space-time, as we extend from the quantum domain into the astronomical domain.

In conclusion, this article illustrates how we can calculate all hadron masses from the Projection-Gell-Mann standard model, and the bare quark mass of 34 MeV with extremely good fits. It also allows us to suggest higher 4, 5 and 6 quarks bound states, and predicting their mass value. Thus experimentally, the theory can be verified conclusively.